\documentstyle[12pt]{article}
\begin{document}
%zzz
\newcommand{\nd}[1]{/\hspace{-0.5em} #1}
\begin{titlepage}
\begin{flushright}
{\bf October 2000} \\ 
SWAT-270  \\ 
hep-th/0010115 \\
\end{flushright}
\begin{centering}
\vspace{.2in}
{\large {\bf Instantons, Compactification and S-duality \\ in 
${\cal N}=4$ SUSY Yang-Mills Theory I}}\\
\vspace{.4in}
 Nick Dorey \\
\vspace{.4in}
Department of Physics, University of Wales Swansea \\
Singleton Park, Swansea, SA2 8PP, UK\\
\vspace{.2in}
%and \\ 
%\vspace{.2in}
%
%
\vspace{.4in}
{\bf Abstract} \\
\end{centering}
We study ${\cal N}=4$ supersymmetric Yang-Mills (SYM) theory with
gauge group $SU(2)$ compactified to three dimensions on a circle of
circumference $\beta$. The eight fermion terms in the effective action
on the Coulomb branch are determined exactly, for all $\beta$, 
assuming the existence of an
interacting $Spin(8)$ invariant fixed point at the origin. 
The resulting formulae are manifestly invariant under the 
$SL(2,Z)$ duality of four dimensional ${\cal N}=4$ SYM 
and lead to interesting quantitative predictions 
for instanton effects in gauge theory and in Type II string theory.

%\vspace{.05in}
%\baselineskip=.3in
\end{titlepage}
\section{Introduction}
\paragraph{}
Gauge theories with extended supersymmetry generically have a
continuous degeneracy of vacua and can be analysed by constructing a Wilsonian
effective action for the corresponding massless fields. 
Typically the lowest non-trivial terms 
in the derivative expansion of the effective action are highly
constrained by supersymmetry and can sometimes be determined exactly. 
In the following we will study ${\cal N}=4$
supersymmetric Yang-Mills theory with gauge group $SU(2)$ compactified
to three dimensions on a circle of circumference $\beta$. 
As in other theories with sixteen supercharges \cite{notes}, 
non-trivial quantum 
effects first appear in terms with four spacetime derivatives. The
supersymmetric completion of the effective action also involves terms
with eight fermions and no spacetime derivatives. In this letter, 
building on previous work in the 
three-dimensional case \cite{PP,DKM97,PSS}, 
we will determine these terms exactly and and extract predictions for
interesting non-perturbative effects in gauge theory and in Type II
string theory. 
\paragraph{}
The approach, based on that 
of \cite{PSS}, is to formulate the condition for unbroken supersymmetry
as a Laplace equation on the vacuum moduli space. The relevant
solution of the Laplace equation is further constrained by its
behaviour near the origin of the moduli space. 
The key assumption here is that the theory at this singular point is an
interacting three-dimensional 
superconformal field theory with $Spin(8)$ R-symmetry\footnote{As
discussed further below, our approach differs slightly from that of
\cite{PSS}. In particular, these authors were able to obtain the exact eight
fermion term in three dimensions {\em without} assuming $Spin(8)$
R-symmetry. Generalizing their derivation to the theory on 
$R^{3}\times S^{1}$ requires some additional technical assumptions. We
will avoid this complication by retaining the assumption of $Spin(8)$
R-symmetry at the conformal point.} \cite{notes,SS,BS}. The symmetries
of this SCFT determine the behaviour of the eight fermion term near
the origin. The exact eight fermion term can then be reconstructed
using elementary facts about the uniqueness of harmonic functions. 
Its notable that the resulting term in the effective action is 
automatically invariant under the 
$SL(2,Z)$ electric-magnetic duality of the four dimensional 
theory. This reinforces the connection between S-duality in
four dimensions and the $Spin(8)$ R-symmetry in three dimensions 
discussed in \cite{notes}. 
\paragraph{}
At weak coupling, 
our result has two distinct instanton expansions, valid near the
three- and four-dimensional limits respectively. 
The model can be realized on the
world volume of two D-branes in Type II string theory on 
$R^{9}\times S^{1}$. In this context, the field theory 
instanton effects exhibit many of the characteristic properties of 
Type II string theory instantons obtained by wrapping D-branes around 
non-trivial cycles in spacetime. In fact, the two expansions have an 
interpretation in terms of wrapped branes of the IIA and IIB 
theories respectively and are related by T-duality. 
In a forthcoming paper \cite{part2}, we investigate these effects via a
direct semiclassical calculation and relate them to the $L^{2}$ index 
theory which counts BPS monopole/dyon boundstates
in four dimensions \cite{sen}. 
\paragraph{}
We begin by considering ${\cal N}=4$ SUSY Yang-Mills theory in four
dimensional Minkowski space with gauge group $G=SU(2)$. 
The theory contains six real scalar fields which
transform in the adjoint representation of the gauge group and the 
${\bf 6}$ of the R-symmetry group $Spin(6)_{\cal R}\equiv SU(4)_{\cal
R}$. Non-zero vacuum expectation values (VEVs) for these 
fields break the gauge group
down to $U(1)$. After taking into account the Higgs mechanism, the
massless fields lie in a single multiplet of ${\cal N}=4$ SUSY which
contains the massless $U(1)$ gauge field $A_{\mu}$, with four
dimensional Lorentz index 
$\mu=0,1,2,3$. We define abelian electric and magnetic fields of the 
low energy theory $E_{i}=F_{0i}=\partial_{0}A_{i}-\partial_{i}A_{0}$ and 
$B_{i}=\varepsilon_{ijk}F^{jk}/2$ with $i=1,2,3$. 
The superpartners of the gauge field include 
neutral Weyl fermions $\lambda^{M}_{\alpha}$ and 
$\bar{\lambda}^{M}_{\dot{\alpha}}$, with $M=1,2,3,4$, in
charge conjugate spinor representations of $Spin(6)_{\cal R}$
as well as neutral scalars $\phi_{a}$, with $a=1,\ldots,6$ in 
the vector representation of this group. We define a $Spin(6)_{\cal R}$
invariant VEV $|\phi|$ with $|\phi|^{2}=\sum_{a=1}^{6} \phi^{2}_{a}$. 
After taking into account the
Weyl group, which acts as $\phi_{a}\rightarrow -\phi_{a}$, the
massless scalars parametrize a classical Coulomb branch which is
$R^{6}/Z_{2}$. 
\paragraph{}
The spectrum of the theory on the Coulomb branch includes BPS states which 
carry electric and magnetic charges with respect to the unbroken
$U(1)$ gauge symmetry. In terms of the low energy fields these are
defined as,    
\begin{equation}
q=\frac{1}{g^{2}} \int d^{3}x \vec{\nabla}\cdot \vec{E}
\label{echarge}
\end{equation}
and, 
\begin{equation}
k=\frac{1}{4\pi} \int d^{3}x \vec{\nabla}\cdot \vec{B}
\label{mcharge}
\end{equation}
respectively. The normalization of the electric charge is chosen so
that a massive vector boson has charge $+1$. The Dirac quantization 
condition then implies
that $k$ is an integer\footnote{Strictly speaking, 
the Dirac quantization condition
only requires $k$ to be half integer. As usual the
consistency of the theory in the presence of charges in the
fundamental representation of the gauge group further restricts $k$ to
integer values}. 
Alternatively, in the full non-abelian theory, the magnetic charge, $k$ is
identified with a non-trivial element of the homotopy group 
$\pi_{2}(SU(2)/U(1))$ which yields the same integer quantization of $k$. 
\paragraph{}
The four dimensional theory has a single parameter, the
complexified coupling $\tau=4\pi i/g^{2}+ \theta/2\pi$ and is believed to be
invariant under $SL(2,Z)$ group of S-duality transformations 
\cite{OM} which act on $\tau$ as $\tau\rightarrow (a\tau+b)/(c\tau+d)$ for 
integers $a,b,c,d$ with $ad-bc=1$. In the
following $M(a,b,c,d)$ denotes the corresponding element of $SL(2,Z)$, 
\begin{eqnarray}
M=\left(\begin{array}{cc} a & b \\ 
c & d \end{array}\right) & \in & SL(2,Z)  
\label{sl2z}
\end{eqnarray}
The spectrum of theory includes BPS states which lie in short
representations of the supersymmetry algebra. 
The states carry the electric and magnetic charges defined above, and 
their masses are determined 
in terms of these quantum numbers to be,   
\begin{equation}
M(q,k)=
|\phi||k\tau+q|=|\phi|\sqrt{k^{2}\left(\frac{4\pi}{g^{2}}\right)^{2}+
\left(q+k\frac{\theta}{2\pi}\right)^{2}}
\label{bps} 
\end{equation} 
The two component charge vector $\vec{q}$ with $q_{1}=q$, $q_{2}=k$
transforms as $\vec{q}\rightarrow \tilde{M}\cdot \vec{q}$ 
under the $SL(2,Z)$ transformation $M(a,b,c,d)$ defined above, where 
$\tilde{M}=M(a,-b,-c,d)$. If we
assign the modular weight\footnote{A modular 
form $f(\tau,\bar{\tau})$ of weight $(w,\bar{w})$ tranforms as 
$f\rightarrow (c\tau + d)^{w}(c\bar{\tau}+d)^{\bar{w}}f$ under the
modular tranformation $T(a,b,c,d)$ discussed in the text.} 
$(1/2,1/2)$ to the scalar fields 
$\phi_{a}$ then the BPS mass formula 
(\ref{bps}) is invariant under S-duality. 
However S-duality of the 
spectrum also requires that a single particle BPS state exists 
whenever $q$ and $k$ are coprime \cite{sen}. 
\paragraph{}
The four dimensional Lorentzian theory with gauge group $SU(2)$ 
can be realized on the world volume of two D3 branes in type IIB
string theory on flat $R^{9,1}$. 
Moving onto the Coulomb branch parametrized by the
six scalar fields of the theory 
corresponds to seperating the branes in their six
transverse directions.  The distance between the D3 branes is proportional
to the $Spin(6)_{\cal R}$ invariant VEV $|\phi|$ 
(in units of $\alpha'$: we will 
henceforth set the string length scale $\sqrt{\alpha'}$ to unity). 
The IIB theory has a complex coupling $\tau$ which is
identified with the complexified gauge coupling of the ${\cal N}=4$ 
theory and an S-duality group which corresponds to the $SL(2,Z)$
invariance of the ${\cal N}=4$ theory on the D3 brane world volume. 
The IIB theory contains $(q,k)$-strings with tensions 
proportional to $|k\tau+q|$ for all coprime values of $q$ and $k$.  
The BPS states discussed above are realized in string
theory as configurations where a $(q,k)$-string is 
stretched between the two D3 branes. The first equality in 
(\ref{bps}) shows that the mass of each BPS state is just the 
tension of the corresponding string multiplied by the distance
between the two branes. 
\paragraph{}
We now switch our attention to the Euclidean theory  
and also compactify 
the Euclidean time direction, $x_{0}$, 
on a circle of circumference $\beta$. 
This breaks the four dimensional Euclidean group to its three-dimensional
counterpart. The $U(1)$ gauge field in four dimensions splits up into
a scalar $A_{0}$ as well as a
three dimensional gauge field $A_{i}$ with Lorentz index $i=1,2,3$. 
The massless modes in three
dimensions correspond to the Fourier modes of the fields with
zero momentum in the compactified direction. 
Physically inequivalent configurations are defined up to
gauge transformations $A_{\mu} \rightarrow A_{\mu}+\partial_{\mu}
\chi(x)$. As all the fields in the
underlying non-abelian gauge theory are invariant under the center of
the gauge group, the necessary condition for such a gauge transformation to be
single valued on $R^3\times S^{1}$ is 
$\chi(x_{0}+\beta)=\chi(x_{0})+2\pi n$. If $n\neq 0$ the 
resulting gauge transformation is topologically non-trivial. The
integer $n$ is an element of $\pi_{1}(S^{1})=Z$ which we will call 
the index of the gauge transformation. These
so-called large gauge transformations will play an important role in
the following. 
\paragraph{} 
In addition to the four dimensional scalars $\phi_{a}$, 
gauge-inequivalent vacua are parametrized by VEV of the Wilson line 
$\omega=\int_{S^{1}} A\cdot dx$. The Wilson line is not invariant under the
large gauge transformations discussed above but rather transforms as
$\omega \rightarrow \omega + 2n\pi$. Hence this field should be
thought of as a periodic variable with period $2\pi$. 
As usual we can eliminate the abelian gauge field
$A_{i}$ in favour of a periodic scalar $\sigma$ via a three dimensional duality
transformation. The field $\sigma$ enters as a Lagrange multiplier for
the Bianchi identity and leads to a surface term in the action, 
$S_{\sigma}=i\sigma k$ where $k$ is the magnetic charge defined in 
(\ref{mcharge}) above. As all the remaining terms in the action
depend on $\sigma$ only via its spacetime derivatives, $\sigma$ is
effectively a periodic variable with period $2\pi$. 
The manifold of gauge-inequivalent classical vacua parametrized by
the VEVs of $\phi_{a}$, $\omega$ and $\sigma$ is  
${\cal M}_{cl}=(R^{6}\times T^{2})/Z_{2}$. 
\paragraph{}
Although the unbroken 
$R$-symmetry of the Lagrangian is just the $Spin(6)_{\cal R}$ of the
four-dimensional theory, the resulting supersymmetry algebra with
sixteen supercharges in three
dimensions\footnote{This is sometimes refered to as $N=8$ SUSY in
three dimensions. Here $N$ counts the number of two-component 
Majorana supercharges.
We prefer to adopt the four-dimensional convention and count the
number ${\cal N}$ of complex two-component 
supercharges which is equal to four in this case.} has a larger
$Spin(8)_{\cal R}$ group of automorphisms. To make this manifest, we
combine the eight scalar fields in a vector with components
$X_{l}$, with $l=1,\ldots ,8$ as: 
\begin{eqnarray}
{\vec X} & = &
\left(\sqrt{\delta}\phi_{1},\ldots,\sqrt{\delta}\phi_{6},
\sqrt{\gamma} \omega, \sqrt{\epsilon}\left(\sigma +
\frac{\theta\omega}{2\pi} \right) \right) \nonumber \\ 
\label{xdef}
\end{eqnarray}
with $\delta=\beta/g^{2}$, $\gamma=1/g^{2}\beta$ and
$\epsilon=g^{2}/(16\pi^{2}\beta)$. The enlarged R-symmetry is such that 
$\vec{X}$ transforms in the
eight-dimensional vector representation ${\bf 8}_{V}$ of
$Spin(8)_{\cal R}$.  
In the
following Roman indices $l,m,n,o=1,\ldots , 8$ label the components of this
repesentation. The low energy theory also includes eight Majorana fermions
$\psi^{\Omega}_{\alpha}$, with $\Omega=1,\ldots, 8$ and $\alpha=1,2$, 
which comprise one of the
two eight-dimensional spinor representations of 
$Spin(8)_{\cal R}$ denoted ${\bf 8}_{S}$. The SUSY
algebra then takes the form, 
\begin{eqnarray} 
\delta X^{l} & = & -i\epsilon^{\alpha}_{\dot{\Omega}}
\Gamma^{l}_{\dot{\Omega}\Sigma} 
\psi_{\alpha\, \Sigma} \nonumber \\ 
\delta \psi_{\alpha
\Sigma} & = & \epsilon^{\beta}_{\dot{\Omega}}
\Gamma^{l}_{\dot{\Omega}\,\Sigma}\tau^{i}_{\beta\alpha}\partial _{i}X^{l}
\nonumber  
\end{eqnarray}
Here $\Gamma^{l}_{\dot{\Omega}\Sigma}$ represent the $Spin(8)$ Clifford
algebra and the dotted and undotted indices $\dot{\Omega}$ and
$\Sigma$ are indices
of ${\bf 8}_{C}$ and ${\bf 8}_{S}$ representations of $Spin(8)_{\cal
R}$ respectively. The $\tau_{i}$ are the $\gamma$-matrices for the
spacetime Clifford algebra in three dimensions. 
\paragraph{} 
The bosonic part of the classical effective action is \cite{SW3} 
$S_{eff}^{B}+
S_{\sigma}$, with,  
\begin{equation}
S_{eff}^{B}= \int\, d^{3}x \, \frac{1}{2}\, \delta_{lm} \partial_{\mu} X^{l}
\partial^{\mu} X^{m} 
\label{seffb}
\end{equation}
While the fermionic part is, 
\begin{equation}
S_{eff}^{F}= \int\, d^{3}x \, \frac{1}{2}\, \delta_{\Omega\Sigma} 
\psi^{\Omega\Sigma} 
\tau^{i}\partial_{i} \psi^{\Sigma} 
\label{sefff}
\end{equation}
This is the action of a three dimensional supersymmetric 
non-linear $\sigma$-model
whose target manifold is 
${\cal M}_{cl}=(R^{6}\times T^{2})/Z_{2}$ with the
standard flat metric. 
The $\theta$-dependent shift in the definition (\ref{xdef}) of $X_{8}$
reflects the fact that the four dimensional $\theta$-term is
proportional to $\vec{E}\cdot\vec{B}$ and therefore yields a coupling
between the spacetime derivatives $\omega$ and $\sigma$. 
Note that the $Spin(8)_{\cal R}$ symmetry is broken to 
$Spin(6)_{\cal R}$ solely by the
periodic boundary conditions on the bosonic fields $X_{7}$ and $X_{8}$;   
\begin{eqnarray} 
X_{7} & \sim & X_{7} - 2n_{2} \Omega_{2} \nonumber \\ 
X_{8} & \sim & X_{8}-2n_{1} \Omega_{1} -2 \kappa n_{2} \Omega_{2}
\nonumber \\
\label{periodicbc}
\end{eqnarray}
with $2\Omega_{1}= g/2\sqrt{\beta}$, $2\Omega_{2}= 2\pi/g\sqrt{\beta}$
and $\kappa=\theta g^{2}/8\pi^{2}$. The real two dimensional torus
parametrized by $X_{7}$ and $X_{8}$ corresponds to a complex torus $E$ with
complex structure parameter $\tau$ and holomorphic coordinate 
$Z=-i(\tau \omega+ \sigma)$. Note that S-duality simply
corresponds to invariance under modular transformations of the complex
torus $E$ \cite{notes}. If we define a two component vector
$\vec{\sigma}$ with components $\sigma_{1}=\sigma$ and
$\sigma_{2}=\omega$, then the transformation law
$\vec{\sigma}\rightarrow \vec{\sigma}'= \tilde{M}\cdot
\vec{\sigma}$ under $M(a,b,c,d)\in SL(2,Z)$, with
$\tilde{M}=M(a,-b-c,d)$ as above, 
ensures that $Z$ transforms with modular weight $(-1,0)$. In the
special case $\theta=0$, where the torus is rectangular then the
periodic boundary conditions (\ref{periodicbc}) break 
$Spin(8)_{\cal R}$ to $Spin(6)_{\cal R}\times Z_{2}$ where the
$Z_{2}$ factor is precisely the $Z_{2}$ subgroup of $SL(2,Z)$
generated by an electric-magnetic duality transformation \cite{notes}.  
\paragraph{}
The limit where the theory reduces
to a gauge theory in three spacetime dimensions is obtained by taking 
$g^{2}\rightarrow 0$ and $\beta\rightarrow 0$ with the three
dimensional gauge coupling $e^{2}=2\pi g^{2}/\beta$ held fixed. 
To get to a generic Coulomb phase vacuum of the $D=3$ theory we must also 
take the Wilson line to zero, holding the three dimensional scalar
field $\phi_{7}=\omega/\beta$ fixed. In
this limit the period $\Omega_{2}$ diverges while $\Omega_{1}$ stays
fixed. As one period of the torus decompactifies the Coulomb branch
becomes $R^{7}\times S^{1}$ and the manifest R-symmetry group 
enlarges from $Spin(6)_{\cal R}$ to $Spin(7)_{\cal R}$. This is the
theory analysed in \cite{PP,DKM97,PSS}. Taking the alternative 
`strong coupling' limit 
$\beta\rightarrow 0$ with $g^{2}$ held fixed, both periods of the
torus decompactify and the Coulomb branch becomes $R^{8}$ 
with a manifest $Spin(8)_{\cal R}$ symmetry at the classical level.       
\paragraph{}
In a theory with sixteen supercharges, supersymmetry does not allow
any corrections to the classical metric on the Coulomb branch. The
action given in (\ref{seffb}) and (\ref{sefff}) above, is therefore
the exact effective action including terms with at most two
spacetime derivatives or four fermions. 
Note that the Coulomb branch has $Z_{2}$ 
orbifold singularities at the four fixed points of the Weyl group: 
$(X_{7},X_{8})=(0,0)$, $(\Omega_{2},0)$,
$(0,\Omega_{1}+\kappa\Omega_{2})$ and 
$(\Omega_{2},\Omega_{1}+\kappa \Omega_{2})$ with $X_{l}=0$ for 
$l\leq 6$ in each case. The origin is
distinguished from the other three fixed points because only at this
point is the non-abelian gauge symmetry restored. As 
new light degrees of freedom appear at this point, we expect that the
low-energy effective theory described above breaks down. In fact it 
is believed that the quantum theory at the origin is an interacting three
dimensional superconformal field theory with an unbroken 
$Spin(8)_{\cal R}$ \cite{notes,SS,BS}. 
In contrast no new light degrees of freedom at the other
fixed points and the low energy 
description remains valid at these points. Hence, apart from the SCFT
at the origin, the theory in each vacuum on the Coulomb branch is 
free in the IR. 
\paragraph{}
Many features of the compactified theory can be understood via its
realization on the world volume of type II branes. As above we start 
with two D3 branes of the type IIB theory. However as 
we wish to discuss the
Euclidean version of ${\cal N}=4$ SUSY Yang-Mills, we now consider the IIB
theory on Euclidean $R^{10}$. After
compactifying one of the directions along the D3 world volume we can
perform a T-duality to the type IIA theory compactified to nine
dimensions on the dual circle. The D3 branes now become D2 branes
located at a point on the $S^{1}$ factor. Moving the branes apart
along the $S^1$ corresponds to introducing a non-zero Wilson line
$\omega$. The periodic nature of this variable is manifest from this
point of view. 
As the dual photon is intrinsically quantum mechanical, the second
compact direction does not appear directly in weakly coupled string
theory. Instead we can invoke the duality between the Type IIA theory on
$R^{9}\times S^{1}$ and M-theory on $R^9\times T^{2}$. 
The D2 branes
lift to M2 branes in eleven dimensions. The radius of the 
eleventh (or M-) direction is smaller by a power of $g^{2}$ than that of 
the other compact direction. The separation of the branes 
in this direction corresponds to the dual photon \cite{T}. The $SL(2,Z)$
invariance of the T-dual IIB theory and therefore of 
${\cal N}=4$ SYM follows naturally from invariance under modular
transformation of the torus $T^{2}$. This ``geometrization'' of
IIB S-duality was first discovered in \cite{schwarz}. The fact that an
analogous geometrical understanding of the Olive-Montonen duality of the
${\cal N}=4$ theory emerges after toroidal compactification was
anticipated in \cite{Harvey,S}. 
\paragraph{}
The M-theory
perspective also provides a natural explanation for the existence of a
$Spin(8)_{\cal R}$ invariant superconformal fixed point in the strong
coupling limit \cite{BFSS2}. 
This limit is precisely the decompactification limit
for the M-dimension and the $Spin(8)_{\cal R}$ invariance is just 
the Euclidean group of spacetime rotations acting on the eight
dimensions transverse to the M2 branes. At least for gauge group
$SU(N)$ with $N$ large, which corresponds to a large number of M2
branes, the resulting three dimensional SCFT is believed to be 
dual to M-theory compactified on $AdS_{4}\times S^{7}$ via the AdS/CFT
correspondence \cite{adscft}. Both the three dimensional conformal group and 
$Spin(8)_{\cal R}$ are manifest as isometries of this spacetime. 
Finally, the appearance of a $Spin(8)_{\cal R}$ invariant fixed point
also plays an important role in the Matrix model of M-theory
\cite{BFSS}. In this
context, it is necessary for an understanding of the limit of 
M-theory on $T^{2}$ which yields IIB string theory on $R^{10}$ \cite{SS}. 
\paragraph{}
As mentioned above, the two derivative terms in the effective 
action are classically exact. 
Nontrivial quantum corrections
to the Wilsonian effective action first appear at the next order 
in the derivative expansion which includes terms with four spacetime
derivatives. The supersymmetric completion of these bosonic terms
include terms with eight fermions and no spacetime derivatives and our
main aim below will be to determine these terms exactly. These
terms are generated in perturbation theory and also recieve instanton
corrections which we will now discuss. The four dimensional
theory contains classical BPS monopoles of unit magnetic charge  
which appear as static solitons of mass $M(0,1)=(4\pi/g^{2})|\phi|$.  
For each value of the magnetic charge $k$, the theory has a 
$4k$-parameter family of exact multi-monopole solutions of mass 
$M(0,k)=kM(0,1)$. After compactification these configuration 
become instantons of 
finite action Euclidean action $\beta kM(0,1)$. Taking into account the
magnetic surface term $S_{\sigma}$ discussed above, these instantons 
yield corrections proportional to $\exp(-\beta kM(0,1)+ ik\sigma)$ to the
low-energy effective action. We will analyse these corrections, and
other related effects in detail in the following and in \cite{part2}.         \paragraph{}
Static BPS monopoles in four dimensions are invariant under half the
supersymmetry algebra. It is often convenient to work in a formalism
where this invariance is manifest. This is accomplished by first
performing an $SU(4)_{\cal R}$ rotation so that only one component of
the Higgs field, say $\phi_{1}$, is non-zero. After choosing
the gauge $A_{0}=0$, we then define an auxiliary Euclidean 
four-dimensional gauge theory with an $SU(2)$ gauge field 
$v_{\tilde{\mu}}$ with $v_{\tilde{\mu}}=A_{i}$ for 
$\tilde{\mu}=i=1,2,3$ and $v_{4}=\phi_{1}$. The easiest
way to motivate this is to recall that  ${\cal N}=4$ SUSY 
Yang-Mills theory can be derived by dimensional reduction of the 
minimal supersymmetric gauge theory in ten dimensions. If our
original four dimensional theory is embedded in the $x_{0}$, 
$x_{i}$ dimensions with $i=1,2,3$ then the auxiliary Euclidean 
theory is simply the corresponding copy ${\cal N}=4$ SYM embedded 
in the $x_{i}$, $x_{4}$ directions. 
Clearly, classical solutions 
which depend only on $x_{i}$ can be thought of as static
configurations in either theory. In the original theory, ten
dimensional Lorentz invariance is broken to the product of the 
four dimensional $SU(2)_{L}\times SU(2)_{R}$ Lorentz group and
$Spin(6)_{\cal R}$. The symmetries of the auxiliary theory 
correspond to a different embedding of these groups into 
$Spin(10)$. We denote the corresponding subgroups, 
$SU(2)_{\tilde{L}}\times SU(2)_{\tilde{R}}\times 
Spin(6)_{\tilde{\cal R}}$. The auxiliary theory contains four 
Weyl fermions of each chirality transforming as 
$({\bf 2},{\bf 0},{\bf 4})$ and $({\bf 0}, {\bf 2},\bar{\bf 4})$ 
respectively under these symmetries. We denote these 
$\rho^{A}_{\delta}$ and $\bar{\rho}^{A}_{\dot{\delta}}$ 
with corresponding indices 
$\delta=1,2$ and $\dot{\delta}=\dot{1},\dot{2}$ and $A=1,2,3,4$.          
\paragraph{}
The advantage of the above construction is that the original
Bogomol'nyi equation for $A_{i}$ and $\phi$ can be rewitten as 
the self-dual Yang-Mills equation for the auxiliary gauge field. 
Thus the magnetic instanton is manifestly invariant under
supercharges of one four-dimensional chirality. 
We emphasize that this refers to the chirality of Weyl spinors in 
the auxiliary theory described above and not in the original
four-dimensional theory. The action of the remaining
supercharges generates eight fermion zero modes of the 
instanton. The Callias index theorem \cite{cal} indicates that the instanton
solution has many additional zero modes. However, as discussed in 
\cite{DKM97}, all of these are lifted by couplings to the scalar
fields of the theory. The instanton therefore has eight exact fermion 
zero modes of the same chirality it contributes to an eight fermion
term in the action built only out of the right-handed Weyl fermions 
$\bar{\rho}^{A}_{\dot{\delta}}$. As these fields have only eight
independent components the form of the vertex is uniquely determined 
to be product of these eight and can be written as,  
\begin{equation}
{\cal L}_{8f}= {\cal V}(\sigma, \omega, |\phi|)\, \prod_{A=1}^{4}\, 
\bar{\rho}^{A}_{\dot{\delta}}\bar{\rho}^{A\,\dot{\delta}} 
\label{eightf2}
\end{equation}
The vertex (\ref{eightf2}) 
contributes to the large distance behaviour of the eight fermion 
correlation function, 
\begin{equation}
{\cal G}^{(8)}(x_{1},\dots,x_{8})=\langle 
\prod_{A=1}^{4} \rho^{A}_{1}(x_{2A}) \rho^{A}_{2}(x_{2A-1}) \rangle 
\label{cor}
\end{equation}  
Strictly speaking, as we started our discussion of the auxiliary
theory by setting $A_{0}=0$, the above result only applies when 
the Wilson line $\omega$ is set to zero. A more complete discussion of
the instanton contribution will be given in \cite{part2}. 
Its also important to note that the chirality selection rule 
apparant in (\ref{eightf2}) is only valid at
leading semiclassical order in a static background. 
In particular perturbative corrections in
the instanton background can (and do) lead to other eight fermion
structures which mix auxiliary fermions $\rho^{A}_{\delta}$ and 
$\bar{\rho}^{A}_{\dot{\delta}}$ of both chiralities. This simply
reflects the fact that the auxiliary four dimensional Lorentz group     
$SU(2)_{\tilde{L}}\times SU(2)_{\tilde{R}}$ is not an exact 
symmetry of the theory. 
\paragraph{}
We will now give an exact analysis the eight fermion terms in the 
effective action. 
To begin with we write down the most general eight fermion term
consistent with the symmetries of the theory. 
The various eight fermion stuctures which can arise can be organised
according to their transformation properties under $Spin(6)_{\cal
R}$. As above we can define complex linear combinations of the 
three-dimensional Majorana fermions which correspond to Weyl fermions 
in four dimensions. These combinations 
transform in the ${\bf 4}$ and $\bar{\bf 4}$ of $Spin(6)_{\cal 
R}\simeq SU(4)_{\cal R}$. 
In general, we must expand all possible products 
of eight fermions, each of which can be either 
in the ${\bf 4}$ or $\bar{\bf 4}$ of $Spin(6)_{\cal R}$, 
as a sum of irreducible representations of this group. 
As this is an unbroken symmetry of the theory we must then form
$Spin(6)_{\cal R}$ invariants by contracting these tensors with 
products of the scalar fields $\phi_{a}$, each of which 
carry a vector index of $Spin(6)_{\cal R}$. Hence for each integer $L$, 
we need only consider eight fermion structures
which transform as rank-$L$ symmetric traceless tensors 
which we denote $T^{(L)}_{a_{1}\ldots a_{L}}(\psi)$. We write the most
general possible coupling as,
\begin{equation} 
{\cal L}_{8f}=\sum_{L} {\cal F}^{(L)}(\vec{X})
X_{a_{1}}\ldots X_{a_{L}}\, T^{(L)}_{a_{1}\ldots a_{L}}(\psi)
\label{so6reps}
\end{equation}
where $X_{a}$ denotes the components of the 
$Spin(8)_{\cal R}$ vector $X_{l}$
introduced above with $l=a=1,\ldots 6$. In the following we will see
that the maximum value of $L$ which contributes to the sum is four.   
The above expansion is complicated by the fact that several linearly 
independent tensor structures may appear for each value of $L$, 
each of which can
appear multiplied by a distinct scalar function of $\vec{X}$. 
Fortunately, our analysis will not be affected by this subtlety and we
have supressed it in (\ref{so6reps}). 
\paragraph{}
The supersymmetric variation of the bosonic fields appearing in the
coefficients of the eight fermion terms leads to terms with nine
fermions. Because all other terms appearing in the 
supersymmetric completion of the four derivative terms have fewer
than eight fermions to start with, there is no other source for 
nine fermion terms in $\delta{\cal L}$. It follows that the nine
fermion terms $\delta {\cal L}_{8f}$ must vanish and this constrains 
the allowed functions ${\cal F}^{(L)}(\vec{X})$.  This constraint was 
studied in detail in \cite{PSS} and we will use their result which
adapts easily to the present case. The result is simply that the eight
fermion Lagrangian ${\cal L}_{8f}$ must be a harmonic function on the 
the Coulomb branch parameterized by $\vec X$. In other words we must
have $\Delta ({\cal L}_{8f})$=0 where $\Delta=\sum_{l=1}^{8} 
\partial^{2}/\partial X_{l}^{2}$ is the Laplacian on ${\cal
M}_{cl}=(R^{6}\times T^{2})/Z_{2}$.
\paragraph{}
The key idea in the following will be the fact that harmonic
functions, like holomorphic functions, are essentially determined by
their behaviour at singularities (and at infinity). Hence if we can
understand the behaviour of the effective action near its singular
points we may reconstruct it everywhere on the moduli
space. In this sense the
argument to follow is a direct generalization of the arguments leading
to the holomorphic superpotential for the ${\cal N}=1^{*}$ theory 
derived in \cite{D}. In fact, when the Coulomb branch is 
two dimensional as it was in that case, a harmonic function is simply
the real part of a holomorphic function.        
\paragraph{}
As usual, singularities may appear at points where 
additional degrees of freedom become light. As mentioned above, it is
believed that the only such point is the origin $\vec{X}=0$ where the 
non-abelian gauge symmetry is restored and the theory becomes
superconformally invariant \cite{notes}. In the following we will 
assume this is 
true and use it to constrain the exact eight fermion terms in the
action. The existence of an interacting SCFT at the origin implies that the
effective action near the origin of the Coulomb branch 
should be invariant under scale transformations and
$Spin(8)_{\cal R}$ rotations. The latter assumption greatly
simplifies the analysis of the eight fermion term.  
As noted above, our approach differs from the
analysis of the three dimensional theory presented in \cite{PSS}. In
particular they did not assume the presence of a $Spin(8)_{\cal R}$
invariant fixed point at the origin but rather proved this directly
from weaker starting assumptions. However, the situation in the compactified
four dimensional theory is more complicated because the unbroken 
R-symmetry on the Coulomb branch is only $Spin(6)_{\cal R}$ compared to the 
$Spin(7)_{\cal R}$ which appears in the three-dimensional limit. It
seems possible that our approach can be adapted to yield the same
results from weaker assumptions but we will not pursue this here. 
\paragraph{} 
In the limit $\vec{X}\rightarrow 0$, we may expand
the product of eight fermions in symmetric, traceless rank-$M$ tensor
representations of $Spin(8)_{\cal R}$, which we denote 
$U^{(M)}_{l_{1}\ldots l_{M}}(\psi)$ and write the 
most general 
$Spin(8)_{\cal R}$ invariant vertex, 
\begin{equation} 
{\cal L}_{8f} \rightarrow \sum_{M} {\cal G}^{(M)}(|\vec{X}|)
X_{l_{1}}\ldots X_{l_{M}}\, U^{(M)}_{l_{1}\ldots l_{M}}(\psi)
\label{so8reps}
\end{equation}
As above there can, in principle, be more than one Lorentz scalar, eight
fermion tensor structure for each $M$. 
The Laplace equation $\Delta {\cal L}_{8f}=0$, decomposes into
decoupled linear equations for each function ${\cal G}^{M}(|\vec{X}|)$
each of which has a unique $Spin(8)_{\cal R}$ invariant solution 
proportional to $|\vec{X}|^{-6-2M}$. Finally we must consider the
restrictions placed by scale invariance. The fields $\vec{X}$ and
$\psi$, lie in a chiral primary multiplet of the superconformal
algebra and their classical scaling dimensions are not corrected by quantum
effects. Thus we have $[X]=1/2$ and $[\psi]=1$ and the action 
is only scale invariant if $[{\cal G}^{(M)}]=-(M+10)/2$. On the other
hand the solutions of the Laplace equation found above have dimension
$-3-M$. Hence scale invariance uniquely selects the $M=4$ term. 
In fact this is exactly the same eight fermion term which appears in
the three dimensional analysis of \cite{PSS}.  
\paragraph{}
One can also check directly that the
instanton induced vertex (\ref{eightf2}), corresponds to 
part of the tensor $U^{(4)}_{l_{1}\ldots l_{4}}$ and has no overlap
with the $M<4$ terms. To see this we 
note that the $Spin(8)_{\cal R}$ symmetry of the conformal point 
includes as a subgroup the $Spin(6)_{\tilde{\cal R}}$ symmetry of the
auxiliary four dimensional theory discussed above. Clearly 
$Spin(8)_{\cal R}$ also includes an additional abelian symmetry, which
we will denote $U(1)_{N}$, that commutes with 
$Spin(6)_{\tilde{\cal R}}$. 
The Weyl fermions of the auxiliary theory, 
$\rho_{\delta}^{a}$ and $\bar{\rho}_{\dot{\delta}}$, are formed from
complex linear combinations of the three-dimensional Majorana 
fermions $\psi^{a}_{\alpha}$. These fields transform 
 $(+1/2,{\bf 4})$ and $(-1/2,\bar{\bf 4})$ under
$U(1)_{N} \times Spin(6)_{\tilde{\cal R}}$ respectively. Hence, the
product of eight right-handed fermions appearing in the instanton
induced vertex (\ref{eightf2}) has the minimum possible 
$U(1)_{N}$ charge $Q_{N}=-4$.      
Its easy to check that the rank-$M$
symmetric traceless tensor representation of $Spin(8)_{\cal R}$
contains states of $U(1)_{N}$ charge $-M \leq Q_{N} \leq +M$. Thus the
maximum value of $M$ which can appear in (\ref{so8reps}) is $M=4$ 
and moreover the instanton vertex (\ref{eightf2}) must correspond 
to part of this structure. 
\paragraph{}
In summary the exact behaviour of eight fermion term in the 
$\vec{X}\rightarrow 0$ limit is,  
\begin{equation} 
{\cal L}_{8f} \rightarrow  
\frac{X_{l_{1}}\ldots X_{l_{4}}}{|\vec{X}|^{14}}\, 
U^{(4)}_{l_{1}\ldots l_{4}}(\psi)
\label{so8reps2}
\end{equation}
This expression is invariant under scale and $Spin(8)_{\cal R}$
symmetries and also satisfies the Laplace equation near the origin of 
the Coulomb branch. Away from the origin the $Spin(8)_{\cal R}$
symmetry will be broken to $Spin(6)_{\cal R}$ by the periodic boundary
conditions on $X_{7}$ and $X_{8}$ and scale invariance will be broken
by explicit dependence on the compactification radius. However
supersymmetry still requires that ${\cal L}_{8f}$ satisfies the
Laplace equation on ${\cal M}_{cl}=(R^{6} \times T^{2})/Z_{2}$ and
this will suffice to determine it exactly. A harmonic function on 
$(R^{6} \times T^{2})/Z_{2}$ defines a harmonic function on
$R^{8}/Z_{2}$ which is invariant under the translations
$\vec{X}\rightarrow \vec{\tilde{X}}$ with $\tilde{X}_{l}=X_{l}$ for 
$l\leq 6$ and  
\begin{eqnarray} 
\tilde{X}_{7} & = & X_{7} - 2n_{2} \Omega_{2} \nonumber \\ 
\tilde{X}_{8} & = & X_{8}-2n_{1} \Omega_{1} -2 \kappa n_{2} \Omega_{2}
\nonumber \\
\label{translate}
\end{eqnarray}      
The required harmonic function must have the behaviour (\ref{so8reps2}) near
the point $\vec{X}=0$ and near each of its images under the above 
translations. As we are solving a linear equation, the required
solution may be generated simply by summing (\ref{so8reps2}) over
$n_{1}$ and $n_{2}$ to get,  
\begin{equation} 
{\cal L}_{8f}=\sum_{n_{1},n_{2}=-\infty}^{+\infty} 
\frac{\tilde{X}_{l_{1}}\ldots \tilde{X}_{l_{4}}}{|\vec{\tilde{X}}|^{14}}\, 
U^{(4)}_{l_{1}\ldots l_{4}}(\psi)
\label{exact}
\end{equation}
Of course an important caveat is that this double sum is convergent. 
In the present case, the large power of $|\vec{\tilde{X}}|$ in the
denominator ensures that this is the case. 
As we have specified the leading
behaviour near each singularity, standard theorems \cite{pde} 
show that, provided this
solution exists, it is unique. In the three dimensional 
limit in which $\Omega_{2}\rightarrow\infty$ with $\Omega_{1}$ held
fixed, we immediately reproduce the results of \cite{PSS}. 
\paragraph{}
Comparing the expression (\ref{exact}), with our starting point 
(\ref{so6reps}) we may read off the exact expressions for the
coefficient functions ${\cal F}^{(L)}(\vec{X})$ of each $Spin(6)_{\cal R}$
representation. We will focus on the highest value of $L$ appearing in
this sum, which is $L=4$. We find, 
\begin{equation} 
{\cal F}^{(4)}(\vec{X})= \sum_{n_{1},n_{2}=-\infty}^{+\infty} 
\frac{1}{\left[
(X_{8}-2n_{1} \Omega_{1} -2 \kappa n_{2} \Omega_{2})^{2}+
(X_{7} - 2n_{2} \Omega_{2})^{2}+ A^{2}\right]^{7}} 
\label{exact2}
\end{equation}
with $A^{2}=\sum_{l=1}^{6} X^{2}_{l}=\beta |\phi|^{2}/g^{2}$. 
Eqn (\ref{exact2}) has a nice interpretation in M-theory where the
theory in question is realized on two M2 branes located a distance
proportional to 
$(A^{2}+X_{7}^{2}+X_{8}^{2})$ apart on their eight common transverse
directions. Eqn (\ref{exact2}) can be a thought of as a sum over
pairwise interactions between the two branes and all of 
their periodic images under the translations (\ref{translate}). 
This generalises the
three-dimensional results of \cite{PP}. It can also be written in a
form which is manifestly invariant under the $SL(2,Z)$ duality of the
four-dimensional theory. In particular we have, 
\begin{eqnarray} 
{\cal F}^{(4)}(\vec{X}) & = & \sum_{n_{1},n_{2}=-\infty}^{+\infty} 
\frac{1}{\left[\frac{g^{2}}{16\pi^{2}\beta}
|Z-n_{2}\tau-n_{1}|^{2}+\frac{\beta}{g^{2}}|\phi|^{2}\right]^{7}} 
\label{exact3}
\end{eqnarray}
As above $Z$, $|\phi|$ and $g^{2}$ have modular weights $(-1,0)$,
$(+1/2,+1/2)$ and $(+1,+1)$ respectively. The summand is modular 
if the two component vector $\vec{n}=(n_{1},n_{2})$ transforms as 
$\vec{n}\rightarrow \tilde{M}\cdot \vec{n}$ under 
$M(a,b,c,d)\in SL(2,Z)$ with $\tilde{M}=M(a,-b,-c,d)$ as above. 
Hence modular transformations effectively permute different terms in
the sum over $n_{1}$ and $n_{2}$. Modular invariance of 
${\cal F}^{(4)}$ then follows after summing over $n_{1}$ and $n_{2}$.    
\paragraph{}
In the remainder of the paper we will examine the physical content of
the above result. First we extract the the perturbative
part of the result which comes
from the sector of zero magnetic charge, 
\begin{equation} 
{\cal F}^{(4)}_{pert}=\left(\frac{g^{2}}{\beta}\right)^{6} 
\sum_{n=-\infty}^{+\infty} 
\frac{1}{\left[ \frac{(\omega-2\pi n)^{2}}{\beta^{2}} + 
|\phi|^{2}\right]^{\frac{13}{2}}} 
\label{pert}
\end{equation}
This term is entirely generated at one loop\footnote{The 
presence of $g^{12}$ as a prefactor reflects our choice of
normalization for the fields}. In the three dimensional limit,
$\beta\rightarrow 0$ and $g^{2}\rightarrow 0$ with $e^{2}=2\pi
g^{2}/\beta$ held fixed, only the $n=0$ term in the sum contributes 
and we recover the behaviour $1/|\phi|^{13}$ of the one-loop
correction computed in \cite{PSS}. After Poisson resummation one may
also take a four dimensional limit which yields a one-loop term 
proportional to $1/|\phi|^{12}$, this is part of the supersymmetric 
completion of the one-loop $F_{\mu\nu}^{4}$ term in the effective 
action of ${\cal N}=4$ SUSY Yang-Mills theory derived in \cite{DS}.  
\paragraph{}
It is straightforward to Fourier expand the result (\ref{exact2}) 
in integer powers of $\exp(i\sigma)$ corresponding to sectors of
different magnetic charge. Explicitly we obtain,    
\begin{equation}
{\cal F}^{(4)}(\vec{X})=\sum_{k=-\infty}^{+\infty}
\sum_{n=-\infty}^{+\infty} {\cal F}_{k,n}(|\phi|,\omega)
\exp\left(ik \left(\sigma+n\theta+\frac{\theta\omega}{2\pi}\right)\right)
\label{fourier2}
\end{equation}
In the weak coupling limit, $g^{2}<<1$, the Fourier coefficient
becomes,  
\begin{eqnarray}
{\cal F}_{k,n} & \sim & \beta^{7} g^{-14} 
\frac{k^{13}}{S_{k,n}^{7}}\, \exp(S_{k,n})
\label{dyonexp2}
\end{eqnarray} 
where, 
\begin{equation} 
S_{k,n}=\frac{4\pi k}{g^{2}} 
\sqrt{\beta^{2}|\phi|^2+ |\omega-2 \pi n|^{2}}
\label{stilde2}
\end{equation}
Hence we have a sum of instanton contributions labelled by two
integers $k$ and $n$. 
To start with we note that, in the special case $\omega=0$, the $n=0$ 
terms yield the expected series of magnetic instanton
corrections with exponential supression $\exp(-\beta k M(0,1) +
ik\sigma)$ where $k$ is identified with the magnetic charge. 
Terms with $n \neq 0$ reflect the phenomenon described in \cite{D,LY}: 
the magnetic instanton is not invariant under large gauge 
transformations. In fact the configurations with action
$S_{k,n}$ given in (\ref{stilde2}) above, are each obtained by
acting on the BPS monopole solution with a large gauge transformation
of index $n$. Note that dependence on $\omega$ and $n$ only arises in
the combination $\omega- 2\pi n$, reflecting the fact that the Wilson 
line is shifted by $2\pi n$ under such gauge transformations. These
instanton effects will be studied directly using semiclassical methods
in \cite{part2}. The exact formula (\ref{exact2}) also predicts a 
series of perturbative corrections to (\ref{dyonexp2}) suppressed by
powers of $g^{2}$. Interestingly, this pertubative series truncates at
a finite order in $g^{2}$. 
\paragraph{}
The instantons described above have a simple interpretation in 
terms of wrapped branes in the IIA theory. As
before our low energy theory lives on two 
D2 branes on $R^{9}\times S^{1}$ where the compact
dimension is transverse to the brane world volume. The branes are
seperated on $S^{1}$ by an amount 
$\omega/\beta$ and in one of the non-compact transverse directions by
distance $|\phi|$. 
The IIA theory contains D0 branes with mass $4\pi/g^{2}$ (in units 
with $\sqrt{\alpha'}=1$). The duality between the IIA string theory 
and M-theory, 
requires that any number of D0 branes form a bound state at threshold.
These bound states are identified with the Kaluza-Klein modes of the 
eleven dimensional metric and the number of constituent D0 brane 
corresponds to momentum in the M-direction. As explained above, 
in the present context, the VEV of the 
dual photon is naturally interpreted as the spatial coordinate in the
eleventh direction. Noting that magnetic monopoles of charge $k$ contribute to
the path integral with a phase $\exp(ik\sigma)$ we immediately see
that magnetic charge should be identified with D0 brane
number. Each of the instantons of magnetic charge $\pm k$ described 
in preceeding paragraphs corresponds to a configuration 
where the Euclidean worldline of a boundstate of $k$ D0 branes is
stretched between the two D2 branes. The possible paths between the
two D2s split up into sectors labelled by an integer $n$ which counts
the number of times the D0 boundstate worldline winds round the compact 
dimension. The shortest path in each sector has length 
$\sqrt{\beta^{2}|\phi|^2+ |\omega-2 \pi n|^{2}}$ and the corresponding
configuration therefore has action $S_{k,n}=(4\pi k/g^{2})
\sqrt{\beta^{2}|\phi|^2+ |\omega-2 \pi n|^{2}}$ in agreement with 
(\ref{stilde2}). In view of this correspondence we will refer to the 
instanton expansion (\ref{fourier2},\ref{dyonexp2}) as the IIA expansion. 
A three dimensional limit in field
theory corresponds to decompactifying the $S^{1}$ direction 
in the IIA picture while keeping the distance between the 
D2 branes fixed. Clearly only the configurations with 
winding number $n=0$ contribute in this limit.
\paragraph{}
Its notable that the exact eight fermion vertex also has another
physically interesting expansion. In particular,
${\cal F}_{4}({\vec X})$ is a periodic function of $\omega$ and
$\sigma$ with period $2\pi$ in each variable. Hence it will have a
double Fourier expansion of the form,  
\begin{equation}
{\cal F}^{(4)}(\vec{X})=\sum_{k=-\infty}^{+\infty}
\sum_{q=-\infty}^{+\infty} \tilde{\cal F}_{q,k}(|\phi|)\exp(-iq \omega + ik
\sigma)
\label{fourier1}
\end{equation}
We will refer to this as the IIB expansion for reasons which will
become clear momentarily. 
It is straightforward to extract the
leading behaviour of the coefficients ${\cal F}_{q,k}$ in the regime
where the compactification radius is large and we are 
`near four dimensions'. For $\beta |\phi|>>1$ we find, 
\begin{eqnarray}
\tilde{\cal F}_{q,k} & \sim & \beta^{-5} g^{12} |\phi|^{-12} 
\,(\beta M(q,k))^{\frac{11}{2}}\, \exp(-\beta M(q,k))
\label{dyonexp}
\end{eqnarray}
where $M(q,k)=|\phi||k\tau + q|$ is the mass of a BPS state of the
four dimensional theory with electric and magnetic charges $q$ and $k$
respectively. However, note that the sum in (\ref{fourier1}) is not
restricted to the coprime values of $q$ and $k$ where we expect the
existence of a BPS bound state in four dimensions. The
exact formulae predict a series of corrections to (\ref{dyonexp}) in
inverse powers of $\beta |\phi|$. However, unlike the perturbative 
corrections to the IIA series (\ref{dyonexp2}), these corrections form
an infinite series. 
\paragraph{}
These effects have simple interpretation in
string theory. We start by returning to ${\cal N}=4$ SUSY 
Yang-Mills in four dimensional Minkowski space which lives on
the world volume of two D3 branes of the IIB theory on $R^{9,1}$. 
As above the ${\cal N}=4$ theory has BPS states with coprime electric and
magnetic charges $q$ and $k$ which are realized as $(q,k)$-strings
stretching between the D3 branes.   
We will now go to Euclidean space and compactify the theory on a 
circle of circumference $\beta$ with
supersymmetry preserving boundary conditions. For $\beta|\phi|>>1$, the
compactification radius is large and, in contrast to our earlier
discussion, it is not appropriate to go to the IIA picture via
T-duality. Rather in the IIB picture we
now have $(q,k)$-strings with Euclidean worldsheets of finite area 
$\beta M(q,k)$ which stretch between the two D3s and also wrap the
compact dimension. These configurations have finite action and
contribute as instantons with the same exponential supression
proportional to $\exp(-\beta M(q,k))$ as the terms in 
(\ref{dyonexp}). However, so far we have only discussed coprime values
of $k$ and $q$. In the more general case suppose $k$ and $q$ have 
maximum common divisor $l>1$, then we have $M(q,k)=lM(q/l,k/l)$ and 
we can obtain a configuration with action $\beta M(q,k)$ by wrapping
the worldsheet of a $(q/l,k/l)$ string stretched between the two D3 
branes $l$ times around the compact dimension. The usual rules of 
D-brane instanton calculus \cite{PK} indicate that such configurations should 
still be invariant under half the supersymmetry algebra and thus
contribute to our eight fermion vertex. Note that similar
configurations involving more than one wrapped 
string have additional zero modes which would prevent
them from contributing.    
\paragraph{}
We will now interpret this 
IIB picture in the context of the gauge theory path integral. 
As usual the Euclidean path integral with periodic boundary conditions 
can be interpreted as a trace over the Hilbert space of the
four-dimensional theory with an insertion of $(-1)^{F}$, where $F$ is
the fermion number operator. 
We will apply this interpretation to the path integral formula for 
the correlation function ${\cal G}^{(8)}$ appearing in (\ref{cor}) 
which corresponds an eight fermion vertex of the form (\ref{eightf2}).  
Taking account of surface terms we find that, 
\begin{eqnarray} 
{\cal G}^{(8)}(x_{1},\dots,x_{8}) & = & \langle 
\prod_{A=1}^{4} \rho^{A}_{1}(x_{2A}) \rho^{A}_{2}(x_{2A-1}) \rangle 
\nonumber \\  
& = & {\rm Tr}
\left[(-1)^{F}
\prod_{A=1}^{4} \rho^{A}_{1}(x_{2A}) \rho^{A}_{2}(x_{2A-1})
\exp\left(-\beta H - i\omega Q+ i\sigma K\right)\right]  
\label{8fq}
\end{eqnarray}
where $H$ is the four-dimensional Hamiltonian and $Q$ and
$K$ are the electric and magnetic charge operators
respectively (which have integer eigenvalues $q$ and $k$). 
This quantity can be thought of as a generalized index. 
Unlike a conventional Witten index, the fermionic insertions ensure
that states of non-zero energy contribute to the trace.  
In particular, the BPS states discussed above 
then contribute to the trace (\ref{8fq}) with the exponential 
supression $\exp(-\beta M(q,k)-iq \omega + ik\sigma)$. 
We will evaluate these contributions explicitly in \cite{part2} and
find that the eight fermionic insertions in (\ref{8fq}) 
effectively act as a projection
operator onto the BPS sector of the Hilbert space so that {\em only} 
BPS states contribute to the trace at leading semiclassical order.  
\paragraph{}
The appearance of the electric surface term $-iq\omega$ in the
exponent of (\ref{8fq}) has an
interesting explanation which we will now sketch. 
In the four dimensional theory,  
one is free to make the gauge choice $A_{0}=0$ and in fact 
this is the convenient choice for discussing the semiclassical 
quantization of monopoles and dyons. After compactification this leads to a 
puzzle because it appears that we have lost the fluctuating 
degree of freedom corresponding to the Wilson line in our 
previous approach! This is not
quite correct however because, to make this gauge choice consistently 
in the Hamiltonian we still have to impose Gauss' law as a constraint.
This can be accomplished by introducing a Lagrange multiplier field 
$\omega$ in close analogy to the way $\sigma$ first appears as a 
Lagrange multiplier for the Bianchi identity. In the latter case the 
result is the magnetic surface term $ik \sigma$ considered above, 
in the former case we obtain $-iq\omega$ where $q$ is the electric 
charge (\ref{echarge}). Note that $\omega$ is a free periodic scalar 
field and the period is $2\pi$ because of electric charge
quantization. Hence we identify $\omega$ with the Wilson line we
have denoted by the same letter above. 
One indication that this identification is correct is that it 
renders the exponent in (\ref{fourier1}) invariant under electromagnetic 
duality.  In particular the combination $iq\omega-ik\sigma$  can be written 
in terms of the 
$SL(2,Z)$ vectors $\vec{\sigma}$ and $\vec{q}$ introduced above as 
$\epsilon_{RS} \sigma_{R}q_{S}$ where $R,S=1,2$ and 
$\epsilon_{RS}$ is the invariant symplectic form of $SL(2,Z)$.
\paragraph{}
So far we have considered instanton expansions arising in 
two different limits of the theory. In each case there are
perturbative corrections to the leading instanton contributions and
the instanton series only makes sense when these are small. 
In this sense, the IIA expansion (\ref{dyonexp2})
is valid for $g^{2}<<1$ while the IIB expansion (\ref{dyonexp}) holds for
$\beta|\phi|>>1$. Interestingly, the regimes in which these two series
are valid have some overlap and, in the following, we will 
investigate the relation between them in their regime of
common validity. In fact when we impose both $g^{2}<<1$
{\it and} $\beta|\phi|>>1$, we will also be able to
compare both expansions 
directly with first principles semiclassical calculations \cite{part2}. 
As a preliminary, we need to 
extract a prediction for the coefficient function ${\cal V}$ of the 
instanton induced vertex (\ref{eightf2}). This requires us to find the
relation between the Majorana fermions appearing in our exact eight
fermion vertex and the four dimensional Weyl fermions
$\rho^{A}_{\delta}$ and $\bar{\rho}^{A}_{\dot{\delta}}$ which appear
in (\ref{eightf2}). Following the discussion above, the components of the 
right handed Weyl fermions $\bar{\rho}^{A}_{\dot{\delta}}$ 
are identified with complex combinations of Majorana fermions which
transform with $U(1)_{N}$ charge $-1/2$. The overall normalizations of the
two sets of fermions are related as $\psi \sim \bar{\rho}
\,\sqrt{\beta/g^{2}}$. For $\beta |\phi|>>1$ and $g^{2}<<1$ we have $A>>
X_{7},X_{8}$ and we find,   
\begin{equation}
{\cal V}(\sigma,\omega,|\phi|) = \left(\frac{\beta}{g^{2}}\right)^{4}
\sum_{n_{1},n_{2}=-\infty}^{+\infty} 
\frac{A^{4}} {\left[
(X_{8}-2n_{1} \Omega_{1} -2 \kappa n_{2} \Omega_{2})^{2}+
(X_{7} - 2n_{2} \Omega_{2})^{2}+ A^{2}\right]^{7}} 
\label{exact5} 
\label{calV}
\end{equation}
\paragraph{} 
Perturbative corrections to the IIA instanton series (\ref{dyonexp2}) can be
neglected if $g^{2}<<1$. If we also impose $\beta |\phi|>> 1$, 
the instanton action 
$S_{k,n}$ appearing in the (\ref{dyonexp2}) 
can be approximated as,  
\begin{equation}
S_{k,n}=k\beta M(0,1) + \frac{1}{2} k \Lambda\, 
\frac{(\omega-2\pi n)^{2}}{\beta} + 
{\rm O}\left( \beta^{-3}|\phi|^{-3} \right)
\label{betasmall}
\end{equation} 
where $\Lambda=M(0,1)/|\phi|^{2}$. 
The resulting IIA series for 
${\cal V}$ is,   
\begin{equation}
{\cal V}_{IIA}= \sum_{k=1}^{\infty}\sum_{n=-\infty}^{+\infty} 
{\cal
V}_{k,n} \exp ik \left(\sigma+n\theta+ 
\frac{\theta \omega}{2\pi}\right)
\label{exp2}
\end{equation}
with 
\begin{eqnarray}
{\cal V}_{k,n}  & = & \left(\frac{\beta}{g^{2}}\right)^{9}\, 
\frac{k^{6}}{(\beta M(0,1))^{3}}\, \exp\left(-\beta kM(0,1) 
-\frac{1}{2}k\Lambda\frac{(\omega-2\pi n)^{2}}{\beta} \right) 
\label{sc2}
\end{eqnarray}
As above, only the $n=0$ term survives in the $D=3$ limit. This term
reproduces the $k$ instanton effect calculated in the
three-dimensional theory in \cite{DKM97}. 
\paragraph{}
On the other hand, if $\beta |\phi|>>1$, perturbative corrections 
(in inverse powers of $\beta|\phi|$) to the IIB series (\ref{dyonexp})
can be neglected. 
If we impose $g^{2}<<1$, we may also approximate the BPS 
mass formula as, 
\begin{equation}
\beta M(q,k)=k\beta M(0,1) + 
\frac{\beta \left(q+\frac{\theta}{2\pi}\right)^{2}}{2k \Lambda} + 
{\rm O}\left(g^{6}\right)
\label{gless}
\end{equation}   
The resulting IIB series for ${\cal V}$ is, 
\begin{equation}
{\cal V}_{IIB}= \sum_{k=1}^{\infty}\sum_{q=-\infty}^{+\infty} 
\tilde{\cal V}_{q,k} \exp (ik \sigma-iq \omega)
\label{exp1}
\end{equation}
with
\begin{eqnarray}
\tilde{\cal V}_{q,k}  & = & \left(\frac{\beta}{g^{2}}\right)^{8}\, \beta \, 
\frac{k^{\frac{11}{2}}}{(\beta M(0,1))^{\frac{5}{2}}}\, 
\exp\left(-\beta k
M(0,1) - \frac{\beta\left(q + \frac{\theta}{2\pi}\right)^{2}}{2k
\Lambda}\right) 
\label{sc1}
\end{eqnarray}
Thus we have two alternative series for ${\cal
V}$, both of which are both valid (in the sense that power law corrections
are small) when $\beta|\phi|>>1$ and also $g^{2}<<1$. However, we must
also consider the convergence of the series themselves. 
In fact the sum over winding number $n$ 
in the IIA series (\ref{exp2}) converges rapidly
only when $\beta|\phi|g^{2}<<1$. On the other hand, rapid convergence
of the sum over electric charge $q$ in the IIB series (\ref{exp1}) 
requires $\beta |\phi| g^{2} >> 1$. Hence the two approximate series 
(\ref{exp2}) and (\ref{exp1}) are useful in complimentary regions of 
parameter space. 
\paragraph{}
It is interesting to investigate the physical content of these 
weak coupling predictions. In particular we can compare the 
detailed expressions (\ref{sc2}) and (\ref{sc1}), 
to a semiclassical instanton calculation on $R^{3}\times
S^{1}$, where the collective coordinates of the monopole are treated 
in the moduli space approximation \cite{manton}. 
Details of these calculations will be given in 
\cite{part2}. However it is easy to understand the exponents
appearing in the approximate IIA and IIB series.  
For this purpose it is be sufficient to recall that, 
in addition to its position in three dimensional space, a single 
BPS monopole in the four dimensional theory 
has an additional modulus describing its orientation in the global
part of the unbroken $U(1)$ gauge group. This collective coordinate is
a periodic variable $\chi\in [0,2\pi]$ and its dynamics are described,
in the moduli space approximation, by the effective Lagrangian 
$L_{\chi}=\Lambda \dot{\chi}^{2}/2$. Here the dot denotes differentiation
with respect to time and, as in (\ref{gless}) above, we have 
$\Lambda=M(0,1)/|\phi|^{2}$. The quantity $\Lambda$ is the
monopole's moment of inertia with respect to global gauge rotations. 
The corresponding Hamiltonian is $H_{\chi}= Q^{2}/2\Lambda$, where
$Q$, the canonical momentum conjugate to $\chi$, is identified with the
electric charge. In the semiclassical limit of the ${\cal N}=4$
theory, we may simply proceed by quantizing this system. 
The condition that the resulting
wavefunctions are single valued on $S^{1}$ naturally leads to 
the integer quantization of electric charge.
After continuation to compact Euclidean time, the resulting 
energy eigenstates 
contribute terms of order $\exp (-\beta q^{2}/2\Lambda)$ 
to the partition function ${\rm Tr}(\exp(-\beta H_{\chi}))$. 
We will call
these states `momentum modes' as they are characterised by a
definite integer value of the conjugate momentum $q$.   
In spacetime these states correspond to an infinite tower of 
BPS dyons which carry magnetic charge one and electric charge $q\in
Z$. As $M(q,1)=M(0,1)+q^{2}/2\Lambda+O(g^{6})$ these states, saturate
the Bogomonl'nyi bound up to higher order corrections in
$g^{2}$. Including the shift in the electric charge due to non-zero 
$\theta$ \cite{witten}, we find contributions with the 
characteristic exponential 
supression of the $k=1$ terms in the approximate IIB expansion 
(\ref{sc1}). In \cite{part2} we will interpret the corresponding 
coefficients of these terms, for all electric and magnetic
charges, as the bulk contribution to the $L^{2}$-index theory 
required to count BPS $(q,k)$-dyons 
in the four-dimensional ${\cal N}=4$ theory.          
\paragraph{}
In fact we can also find a corresponding 
weak coupling interpretation for the $k=1$
terms in the approximate IIA series (\ref{sc2}). As above the 
semiclassical contributions are proportional to the 
quantum mechanical partition function 
${\rm Tr}(\exp(-\beta H_{\chi})$. 
Rather than evaluating this trace directly in canonical quantization, 
we may instead consider the corresponding quantum mechanical path 
integral in compact Euclidean time $\tau\in[0,\beta]$, 
\begin{equation}
{\rm Tr}(\exp(-\beta H_{\chi})=\int\, [d\chi(\tau)]
\exp\left(-\int_{0}^{\beta} \, d\tau\, L_{\chi}\right) 
\label{pint}
\end{equation}
As the dimensionless parameter $\Lambda\beta$ is very large at
 weak coupling we may evaluate this path integral in the semiclassical
approximation. This requires us to 
sum over classical paths $\chi(\tau)$ with periodic boundary conditions 
$\chi(\tau+\beta)=\chi(\tau)+2\pi n$ for integer $n$. 
Hence the admissable classical
paths are $\chi(\tau)=2\pi n\tau /\beta$. As these paths describe 
the worldline of a point particle of mass $\Lambda$ winding $n$ times 
around the target circle, we will call these configurations 
`winding modes'.  
The classical action of a winding mode is precisely 
$2\Lambda \pi^{2}n^{2}/\beta$. Adding in
the action of the static monopole $\beta M(0,1)$ we reproduce the
exponential supression of the $k=1$ terms in (\ref{sc2}) 
for the case $\omega=0$. In spacetime, these winding modes correspond  
to dyon-like classical solutions which spin in the internal $U(1)$. 
In particular, they complete exactly $n$ orbits of $U(1)$ 
in the periodic Euclidean time interval $[0,\beta]$.
Finally note that because the collective coordinate $\chi$
parametrises global $U(1)$ gauge transformations, the  
$\tau$-dependence is identical to the $x_{0}$ dependence of a 
large gauge transformation of
winding number $n$ acting on the monopole. This agrees perfectly with 
the identification, given above, of the summation variable 
$n$ appearing in the IIA series as the winding number of a large gauge
transformation. 
\paragraph{}
In fact it is well known that the semiclassical approximation to the
path integral (\ref{pint}) is actually exact \cite{schulman}. 
The resulting sum over winding modes coming from the 
RHS of (\ref{pint}) is equal 
to the sum over momentum modes appearing in the trace on the LHS. 
As a consequence, the the approximate IIA/IIB series (\ref{exp2}) and
(\ref{exp1}) are actually exactly equal: ${\cal V}_{IIA}={\cal
V}_{IIB}$. In particular we have,  
\begin{equation} 
\sum_{n=-\infty}^{+\infty} {\cal V}_{k,n}
\exp\left(ik\theta\frac{(\omega+2\pi n)}{2\pi}\right)= 
 \sum_{q=-\infty}^{+\infty}\tilde{\cal V}_{q,k}\exp(-iq\omega )
\label{equality}
\end{equation}
This equality follows immediately from the Poisson resummation formula, 
\begin{equation}
\sum_{q=-\infty}^{+\infty} \, \exp \left[ -i \pi B q^{2} + 2\pi i s
q\right] = \frac{1}{\sqrt{iB}} \sum_{n=-\infty}^{n=+\infty} \exp
\left[\frac{i \pi}{B} (s+n)^{2} \right] 
\label{poisson}
\end{equation} 
with $B=\beta/2\pi i \Lambda$ and $s=\omega/2\pi$. Of course this 
Poisson-Lie duality between momentum and winding modes is nothing
other than the T-duality between the IIB and IIA pictures described
above.  As expected, the IIB series is 
convergent near four dimensions ($\beta|\phi|g^{2} >>1$) 
and the IIA series is convergent near three dimensions 
($\beta|\phi|g^{2}<<1$). 
\paragraph{}
The author would like to thank Tim Hollowood and Andrei 
Parnachev for useful discussions. The author also acknowledges the
support of a PPARC Advanced Reseach Fellowship.

\end{document}